\documentclass{jltp}
\usepackage{graphicx}
\title{Cascade process of vortex tangle dynamics\\ in superfluid $^4$He without mutual friction}

\def\Vec#1{\mbox{\boldmath $#1$}}
\author{Tsunehiko Araki and Makoto Tsubota}
\address{Department of Physics,Osaka City University,Osaka 558-8585, Japan}

\runninghead{T.Araki and M.Tsubota}{Cascade process of vortex tangle dynamics without mutual friction}

\begin{document}

\maketitle

\begin{abstract}
Recently the Kelvin wave cascade process in superfluid $^4$He at very low temperatures was discussed. In this mechanism, the dynamics of the waves on the vortex lines plays an important role. In order to investigate this mechanism, we study numerically the dynamics of the waves on the reconnected vortex lines using the full Biot-Savart law. This work shows the reconnection of two vortices leads to the waves on the reconnected vortex lines. To discuss the energy of the waves on the vortex lines, the energy spectrum of the superflow before and after the reconnection is calculated.

\end{abstract}

\section{INTRODUCTION}
Davis et al. observed the free decay of a vortex tangle in superfluid $^4$He at mK temperatures\cite{Davis}. The first important point is that the vortex tangle does decay in spite of the absence of the normal fluid and the mutual friction. The second is that the decay rate becomes independent of the temperature below 70mK. This mechanism is unknown. Above 1K the vortex tangle is maintained by the driving flow; if the driving flow is switched off, the normal fluid collides with vortices and takes energy from them. The microscopic picture of the motion of the vortex tangle subject to the mutual friction is revealed numerically by Schwarz\cite{Schwarz}.
 
 In our previous paper, the dynamics of a dense tangle without the mutual friction was studied numerically\cite{Tsubota}. The absence of the mutual friction makes the vortices kinked. Then lots of small vortex rings appear through the self-reconnection of such kinked parts. The resulting vortices also follow the self-similar process breaking up to smaller ones. According to the discussion of the dynamical scaling\cite{Schwarz,Swanson}, this process is expected to continue self-similarly down to a microscopic scale. Eventually the microscopic rings might degenerate to elementary excitations, which is beyond this formulation. This decay mechanism due to the cascade process is nothing but that proposed once by Feynman\cite{Feynman}.

Recently Vinen proposed another decay due to the acoustic emission from an oscillating vortex at very low temperatures\cite{Vinen}. This process can be divided into three stages. The first is that reconnections leave sharp kinks on the vortex lines. The second is that these kinks evolve the Kelvin waves. The vortex line configuration on a length scale smaller than the distance of vortex spacing $l$ may be described by the Kelvin wave. The third is that the Kelvin waves with wavenumber which is larger than $\tilde{k}_{2}$ are strongly damped by the acoustic emission. The first and second stages are discussed also by Svistunov\cite{Svistunov}. Vinen estimates $\tilde{k}_2$ by the dimensional analysis as\cite{Vinen}
\begin{equation}
\tilde{k}_2=\left( \frac{C}{A^{1/2} \kappa l} \right)^{1/2}, \label{eq.1}
\end{equation}
where $C$ is the speed of sound, $A$ a constant and $\kappa$ the quantized circulation. Equation (\ref{eq.1}) shows $\tilde{k}_2$ is much larger than $2\pi/l$.

In the above mechanism, energy flow between waves with wavenumbers in the range $2\pi/l$ to $\tilde{k}_2$ is very important, because it is closely related with the decay rate. Since the above mechanism occurs as the result of non-linear vortex dynamics, it can be confirmed by the numerical analysis, although our numerical model does not include the effect of the acoustic emission. This work shows the reconnection of two vortices leads to the waves on them. To estimate the energy flow of the waves on the vortex lines, the energy spectrum of the superflow is calculated.

\section{NUMERICAL PROCEDURE}

The configuration of vortices is calculated by the numerical method which is very similar to that of Schwarz and described in our previous paper\cite{Tsubota}. In our calculation, a vortex filament is represented by a single string of points at a distance $\Delta \xi_1$ apart; this distance yields the first space resolution in this calculation. We prepare the initial configuration of vortex lines. When two vortices approach within $\Delta \xi_1$, it is assumed that they are reconnected. The superfluid velocity field is determined by the configuration of vortices.

We introduce the energy spectrum of this velocity field\cite{Nore}. The kinetic energy can be defined as the integral of the square of a field:
\begin{equation}
E_{\rm kin}=\frac{1}{2(2\pi)^3}\int d^3 x (\sqrt{\rho}\Vec{v})^2, \label{eq.2}
\end{equation}
where $\rho$ is the density of fluid and $\Vec{v}$ the velocity field. The energy spectrum $E_{\rm kin}(k)$ is defined as $E_{\rm kin}=\int_{0}^{\infty}dk E_{\rm kin}(k)$. Using the Parseval's theorem, one gets the following energy spectrum:
\begin{equation}
E_{\rm kin}(k)=\frac12 \int d\Omega_{k} \left| \frac{1}{(2\pi)^3}\int d^3 r e^{i \Vec{r}\cdot \Vec{k}} \sqrt{\rho}\Vec{v} \right|^2 , \label{eq.3}
\end{equation}
where $d \Omega_{k}$ denotes the volume element $k^2 \sin \theta d\theta d \phi$ in the spherical coordinates. Since the energy spectrum $E_{\rm kin}(k)$ represents the contribution from the velocity field with wave number $k$ to the kinetic energy, $E_{\rm kin}(k)$ reflects the characteristic scale of vortices; e.g., the distance of vortex spacing $l$, the radius of curvature $R$ and the wavelength of the vortex waves. In order to obtain the energy spectrum, we calculated the velocity on 3D fixed lattice points; this lattice constant $\Delta \xi _2$ is the second space resolution.

\section{ENERGY SPECTRUM BEFORE AND AFTER RECONNECTION}

We calculated the collision of a straight vortex line and a moving ring by the full Biot-Savart law in order to study the waves on the reconnected vortex lines. The computation sample $L$ is taken to be a cube of size 1cm. The calculation is made by the space resolution $\Delta \xi_{1}=4.58 \times 10^{-3}$cm and the time resolution $\Delta t_{1}=6.25 \times 10^{-5}$sec. This calculation assumes the walls to be smooth and takes account of image vortices. Figure 1(a) shows the initial configuration of vortex lines.  Toward the reconnection, the ring and the line twist themselves so that they become locally antiparallel at the closest place (Fig.1(b)). In the following, the time is normalized as $\tilde{t}=t/0.625$sec. At $\tilde{t} \simeq 20.4$, two vortices reconnect. After the reconnection (Fig.1(c), (d) and (e)), the resulting local cusps broaden while exciting vortex waves with various wavenumbers (Fig.1(f)). It is found that the reconnection leads to the waves on the vortex lines.

In order to discuss the energy flow in the process of Fig.1, the energy spectrum before and after the reconnection is calculated (Fig.2). This calculation is made by the space resolution $\Delta \xi_2 =3.90 \times 10^{-3}$cm, the time resolution $\Delta t_2 =6.25 \times 10^{-2}$sec and $\tilde{k}$ the wavenumber which is normalized by $2\pi$. Since this calculation does not include the dissipation mechanism, the total kinetic energy is conserved within the numerical error. Figure 2 shows two characteristic results. The first is that, after the reconnection ($\tilde{t} \ge 20.4$), the fluctuation of the energy spectrum is increased suddenly. The waves generated by the cusps evolve chaotically to waves with other wavenumber by the nonlinear interaction; the reconnection process results in the ergodic energy distribution of vortex waves. This is consistent with the study of the sideband instability by Samuels and Donnelly\cite{Samuels}. The second is that some energy peaks move to small $k$ region. Compared with Fig.1, this behavior of the peaks may reflect that broadening of the local cusps and the ring's leaving the line. If some dissipative mechanism work in the $k$ region above a characteristic wavenumber, e.g., $\tilde{k}_2$ in the Vinen's theory, the kinetic energy is transferred from small $k$ region to large one being dissipated at $\tilde{k}=\tilde{k}_2$ by the nonlinear interaction, so that the total kinetic energy decays. This mechanism is consistent with the Vinen's theory. 
\begin{figure}
%
\centerline{\includegraphics[height=3.6in]{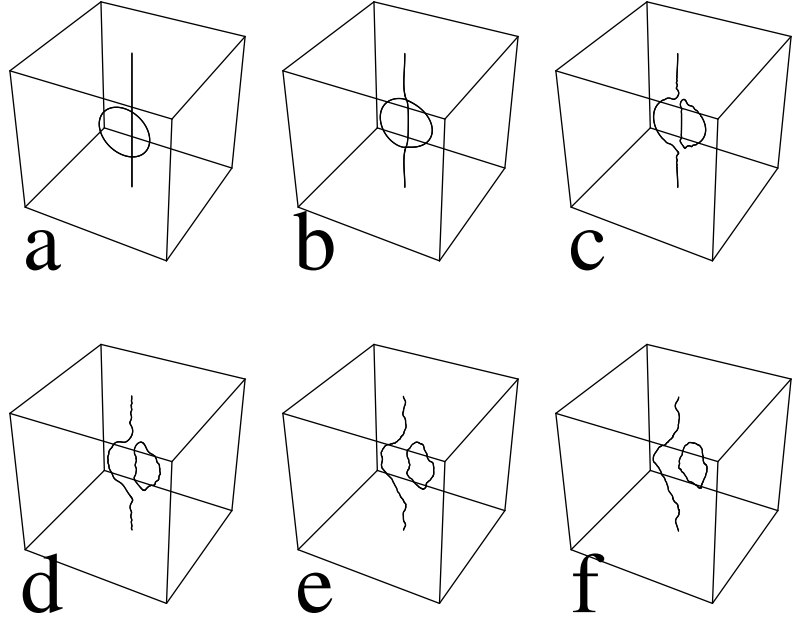}}
%
\caption{Collision of a straight vortex and a ring at (a)$\tilde{t}=0$, (b)$\tilde{t}=20$, (c)$\tilde{t}=21$, (d)$\tilde{t}=22$, (e)$\tilde{t}=24$ and (f)$\tilde{t}=26$. At $\tilde{t} \simeq 20.4$, two vortices reconnect.}
\label{fig:tau2} \end{figure}
\begin{figure}
\centerline{\includegraphics[height=3.85in]{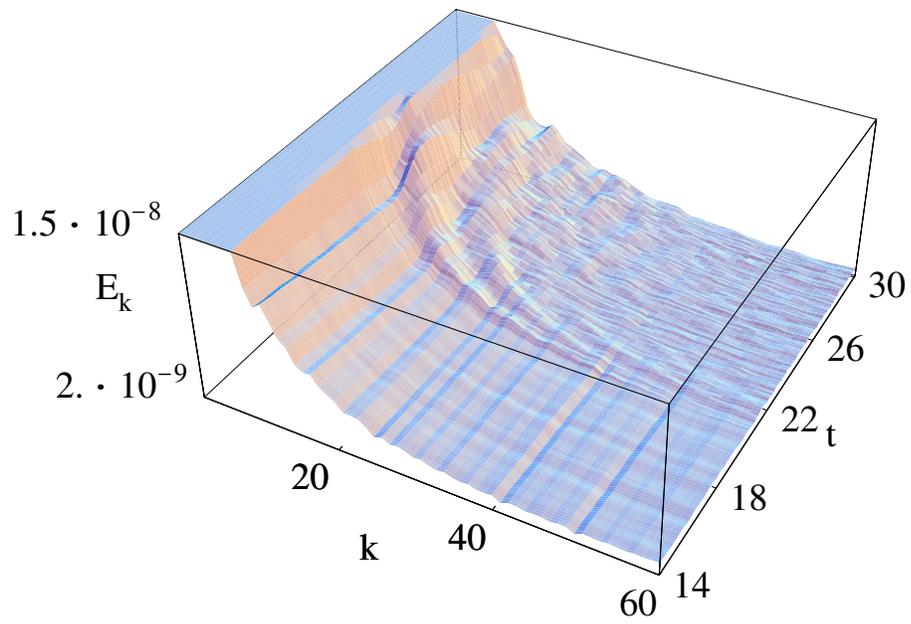}}
%
%
\caption{Time evolution of the energy spectrum $E(\tilde{k})$ before and after the reconnection.}
\label{fig:tau2} \end{figure}
\section{CONCLUSIONS}

This work studies numerically the waves on the reconnected vortex lines. The reconnection process results in the waves on the vortex lines. The ergodic distribution of wavenumbers is generated by the nonlinear interaction of vortex waves. If some dissipative mechanism work in the $k$ region above a characteristic wavenumber, the kinetic energy is transferred from small $k$ region to large by the nonlinear interaction, so that the total kinetic energy decays.

\section*{ACKNOWLEDGMENTS}
We would like to thank W.F.Vinen for many helpful discussions and useful suggestions.

\end{document}